\documentclass[10pt,twocolumn,twoside]{IEEEtran}

\usepackage{amssymb}
\usepackage{amsfonts}
\usepackage{amsmath}
\usepackage{cite}
\usepackage[dvips]{graphicx}
\usepackage{color}
\usepackage{comment}
\usepackage{makecell}
\usepackage{textcomp,mathcomp}
\usepackage{subfigure}
\usepackage{multirow}

\begin{document}
\newtheorem{definition}{\it Definition}
\newtheorem{theorem}{\bf Theorem}
\newtheorem{lemma}{\it Lemma}
\newtheorem{corollary}{\it Corollary}
\newtheorem{remark}{\it Remark}
\newtheorem{example}{\it Example}
\newtheorem{case}{\bf Case Study}
\newtheorem{assumption}{\it Assumption}
\newtheorem{property}{\it Property}
\newtheorem{proposition}{\it Proposition}

\newcommand{\hP}[1]{{\boldsymbol h}_{{#1}{\bullet}}}
\newcommand{\hS}[1]{{\boldsymbol h}_{{\bullet}{#1}}}

\newcommand{\ba}{\boldsymbol{a}}
\newcommand{\baq}{\overline{q}}
\newcommand{\bA}{\boldsymbol{A}}
\newcommand{\bb}{\boldsymbol{b}}
\newcommand{\bB}{\boldsymbol{B}}
\newcommand{\bc}{\boldsymbol{c}}
\newcommand{\bcO}{\boldsymbol{\cal O}}
\newcommand{\bh}{\boldsymbol{h}}
\newcommand{\bH}{\boldsymbol{H}}
\newcommand{\bl}{\boldsymbol{l}}
\newcommand{\bm}{\boldsymbol{m}}
\newcommand{\bn}{\boldsymbol{n}}
\newcommand{\bo}{\boldsymbol{o}}
\newcommand{\bO}{\boldsymbol{O}}
\newcommand{\bp}{\boldsymbol{p}}
\newcommand{\bq}{\boldsymbol{q}}
\newcommand{\bR}{\boldsymbol{R}}
\newcommand{\bs}{\boldsymbol{s}}
\newcommand{\bS}{\boldsymbol{S}}
\newcommand{\bT}{\boldsymbol{T}}
\newcommand{\bw}{\boldsymbol{w}}

\newcommand{\balpha}{\boldsymbol{\alpha}}
\newcommand{\bbeta}{\boldsymbol{\beta}}
\newcommand{\bOmega}{\boldsymbol{\Omega}}
\newcommand{\bTheta}{\boldsymbol{\Theta}}
\newcommand{\bphi}{\boldsymbol{\phi}}
\newcommand{\btheta}{\boldsymbol{\theta}}
\newcommand{\bvarpi}{\boldsymbol{\varpi}}
\newcommand{\bpi}{\boldsymbol{\pi}}
\newcommand{\bpsi}{\boldsymbol{\psi}}
\newcommand{\bxi}{\boldsymbol{\xi}}
\newcommand{\bx}{\boldsymbol{x}}
\newcommand{\by}{\boldsymbol{y}}

\newcommand{\cA}{{\cal A}}
\newcommand{\bcA}{\boldsymbol{\cal A}}
\newcommand{\cB}{{\cal B}}
\newcommand{\cE}{{\cal E}}
\newcommand{\cG}{{\cal G}}
\newcommand{\cH}{{\cal H}}
\newcommand{\bcH}{\boldsymbol {\cal H}}
\newcommand{\cK}{{\cal K}}
\newcommand{\cO}{{\cal O}}
\newcommand{\cR}{{\cal R}}
\newcommand{\cS}{{\cal S}}
\newcommand{\dcS}{\ddot{{\cal S}}}
\newcommand{\ds}{\ddot{{s}}}
\newcommand{\cT}{{\cal T}}
\newcommand{\cU}{{\cal U}}
\newcommand{\wt}[1]{\widetilde{#1}}

\newcommand{\mA}{\mathbb{A}}
\newcommand{\mE}{\mathbb{E}}
\newcommand{\mG}{\mathbb{G}}
\newcommand{\mR}{\mathbb{R}}
\newcommand{\mS}{\mathbb{S}}
\newcommand{\mU}{\mathbb{U}}
\newcommand{\mV}{\mathbb{V}}
\newcommand{\mW}{\mathbb{W}}

\newcommand{\uq}{\underline{q}}
\newcommand{\ubq}{\underline{\boldsymbol q}}

\newcommand{\red}[1]{\textcolor[rgb]{1,0,0}{#1}}
\newcommand{\gre}[1]{\textcolor[rgb]{0,1,0}{#1}}
\newcommand{\blu}[1]{\textcolor[rgb]{0,0,0}{#1}}

\title{Towards Net-Zero Carbon Emissions in Network AI for 6G and Beyond} 

\author{Peng~Zhang, Yong~Xiao, Yingyu Li, Xiaohu~Ge, Guangming~Shi, and Yang~Yang 

\thanks{
This work has been accepted at IEEE Communications Magazine (URL: https://doi.org/10.1109/MCOM.003.2300175). Copyright may be transferred without notice, after which this version
may no longer be accessible. 

{\it (Corresponding author: Yong Xiao)}

Peng Zhang, Yong Xiao, and Xiaohu Ge are with the Huazhong University of Science and Technology; 
Yong Xiao is also with Peng Cheng Laboratory and Pazhou Lab (Huangpu); 
Yingyu Li is with the School of Mechanical Engineering and Electronic Information, China University of Geosciences (Wuhan); 
Guangming Shi is with the Peng Cheng Laboratory and also 
with the Xidian University and also Pazhou Lab (Huangpu);
%
Yang Yang is with the Hong Kong University of Science and Technology (Guangzhou). 
%



%
}
}
\maketitle
\begin{abstract}
A global effort has been initiated to reduce the worldwide  greenhouse gas (GHG) emissions, primarily carbon emissions, by half by 2030 and reach net-zero by 2050. The development of 6G must also be compliant with this goal. Unfortunately, 
developing a sustainable and net-zero emission systems to meet the users' fast growing demands on mobile services, especially smart services and applications, may be much more challenging than expected. Particularly, despite the energy efficiency improvement in both hardware and software designs, the overall energy consumption and carbon emission of mobile networks are still increasing at a tremendous speed. The growing penetration of resource-demanding AI algorithms and solutions further exacerbate this challenge. In this article, 
we identify the major emission sources and introduce an evaluation framework for analyzing the lifecycle of network AI implementations. A novel joint dynamic energy trading and task allocation optimization framework, called DETA, has been introduced to reduce the overall carbon emissions. We consider a federated edge intelligence-based network AI system as a case study to verify the effectiveness of our proposed solution. Experimental results based on a hardware prototype suggest that our proposed solution can reduce carbon emissions of network AI systems by up to 74.9\%. Finally, open problems and future directions are discussed. 
\end{abstract}

\section{Introduction}
\label{Section_Introduction}

Recently, there is a growing interest in the climate change and energy sustainability for ICT industry\cite{Arora2019UNGoals}. Previous studies have already suggested that, 
to preserve a livable planet and avoid the worst-case impacts of climate change, the global 
greenhouse gas (GHG) emissions, primarily carbon emissions, must be reduced by at least 45\% by 2030 and reach net-zero by 2050. To achieve this goal, a worldwide effort has been initiated with more than 70 major countries, including China, United States, and European Union, announced plans to reach the net-zero carbon emission target. 
Furthermore, over 1,200 companies, 1000 educational institutions, and 400 financial institutions, across more than 1000 cities have already pledged to take rigorous and immediate actions to reduce global carbon emissions by half by 2030. 



The future evolution of mobile technology, including  6G and beyond, must also be compliant with the goal of 
net-zero carbon emissions society in 2030 and beyond. Unfortunately, 
the recently published vision and demand for 6G
pose unprecedented challenges for the ICT industry to meet the above goal. 
Recent study suggests that the continuous improvement of communication and networking performance does not come at no cost.
For example, compared to 4G, the improved performance offered by 5G results in around 70\% increase in power consumption of the network infrastructure. With the continuous deployment  
of high-performance networking systems such as 5G-Advanced, industrial IoT, and connected vehicles, 
the total amount of energy consumption of the ICT industry may exceed 20\% of the global energy supply in 2030. 


Another critical challenge faced by 6G is its continuous integration with AI\cite{Dang20206GNatureEle, XY2020Selflearning, YangYang20226G}. Recent study has reported that the training cost of the state-of-the-art AI algorithms, the deep learning algorithms in particular, has increased 300,000 times from 2012 to 2017\cite{Schwartz2020GreenAI}. In other words, the current trends in AI research and model development are both environmentally unfriendly and prohibitively expensive. Another study published by NVIDIA claimed that the training phrase only accounts for around 10-20\% of total energy consumption of an AI algorithm, while most of the energy is spent on model serving phrase. In other words, the increasing penetration of AI algorithms and solutions throughout the already resource-demanding networking systems 
may eventually jeopardize the global initiative towards the sustainable growth and net-zero emission human society in 2030 and beyond\cite{Stefanos2022GreenAI}. Therefore, there is a pressing need to develop novel solutions to evaluate and optimize the environmental impact of the potential AI implementations throughout various parts of networking systems. 

Despite its urgency, the carbon emissions of AI and networking convergence is relatively under-explored in the literature.  
In this article, we 
introduce an evaluation framework for analyzing the lifecycle of a network AI implementation and identify the key sources of carbon emissions of network AI-based systems. A novel optimization solution, called dynamic energy trading and task allocations (DETA), has been introduced to optimize the carbon emission of the entire lifecycle of any given network AI implementation. We 
consider a network AI implementation in a federated edge intelligence (FEI)-based network as a case study. 
Experimental results suggest that the proposed optimization solution has the potential to significantly reduce the carbon emission of the future networking systems. 

\section{Life Cycle Assessment of Network AI}
\label{Section_CarbonEmissionNetworkAI}


According to ITU-T, the carbon emissions of a networking device can be categorized into three classes: {\it Scope 1 Emission:} corresponds to the carbon dioxide emitted directly from the devices themselves. {\it Scope 2 Emission: } corresponds to the carbon dioxide (CO$_2$) emission caused by the generation of electricity that powers the devices. and {\it Scope 3 Emission: } corresponds to other indirect carbon dioxide emission generated by various parts of the networks' or networking devices' value chain\cite{ITUCarbonEmissionScope}. Scope 1 emissions include carbon emissions caused by the manufacturing processes and that generated by burning fuel in vehicles, generators, or furnaces, and also scope 3 emissions are closely related to the supply chain, both of which have limited correlation to the computational and communication networking process in 6G and are therefore out of the scope of this article. It is known that the carbon emissions of the existing networking systems are dominated by the scope 2 emissions.   Therefore, we mainly focus on the scope 2 emission in the rest of this article.

We propose a novel evaluation framework to assess the lifecycle of a network AI implementation from the perspective of model development and implementation pipeline. In particular, the lifecycle of an AI solution implementation can be roughly divided into the following four stages:

\noindent
{\bf Preparation stage:} This includes  collection and transportation of data from the data collecting devices, e.g., IoT devices, to the storage and processing servers, e.g., cloud/edge server. It also involves the identification of specific service needs and requirements of AI models. In particular, before implementing any specific algorithms and solutions, the collected data should be pre-processed according to the requirements and configurations of the available  software and hardware platform as well as the service and application scenarios. The carbon emission in this stage is therefore mainly related to the energy production and consumption of the data generating devices, e.g., IoT devices, data transportation networks, e.g., RAN and backbone networks, and the data storage and/or preprocessing servers at the cloud and edge computing servers.

\noindent
{\bf Development stage:} This includes the model design, algorithm development, as well as parameter selection and optimization. It also involves model pre-training, testing, as well as simulating and adjusting according to the specified service requirements and use scenarios. The carbon emission in this stage is primarily related to the process of energy generation, delivery, and consumption of edge servers and cloud data center.  The carbon emission in the development stage is generally difficult to keep track of because it can be closely related to the experience and skillsets of the model development team as well as the software and hardware platforms that are available at different time of the development stage. Previous study has shown that, compared to the development stage, for most companies, the application stage consumes much more energy and generates more carbon emissions, especially for popular AI models. Recently, the development of Large language models (LLMs) such as ChatGPT has attracted significant interest. These models consume much more energy in the development stage, compared to the previous models. For example, GPT-3, an earlier version of ChatGPT, has already consumed 1,287 MWh of electricity which is equivalent to 552 kg CO$_2$e, for model training\cite{David2021NeuralNetworkCarbon}. 


\noindent
{\bf Application stage:} This stage involves the model distribution, implementation, serving, and possible interaction under the specific platform and use scenarios. 

\noindent
{\bf Recycling stage:} Unlike the hardware devices, most AI implementations may not have a clear end-of-life time, due to the fact that most of these algorithmic implementations can be modified for some other use scenarios and applications in the future. 
Also, their accumulated experience and knowledge will  be helpful for improving the efficiency and performance of other AI implementations in the future. We  therefore define the final stage of AI implementations  as the recycling stage which includes the generalization, modification, and transferring of AI algorithms and implementations according to future environments, configurations, as well as applications and services.


\section{Carbon Emission Sources of 6G Network AI}
\label{Section_Carbon6GNet}
\begin{figure}
\centering
\includegraphics[width=9cm]{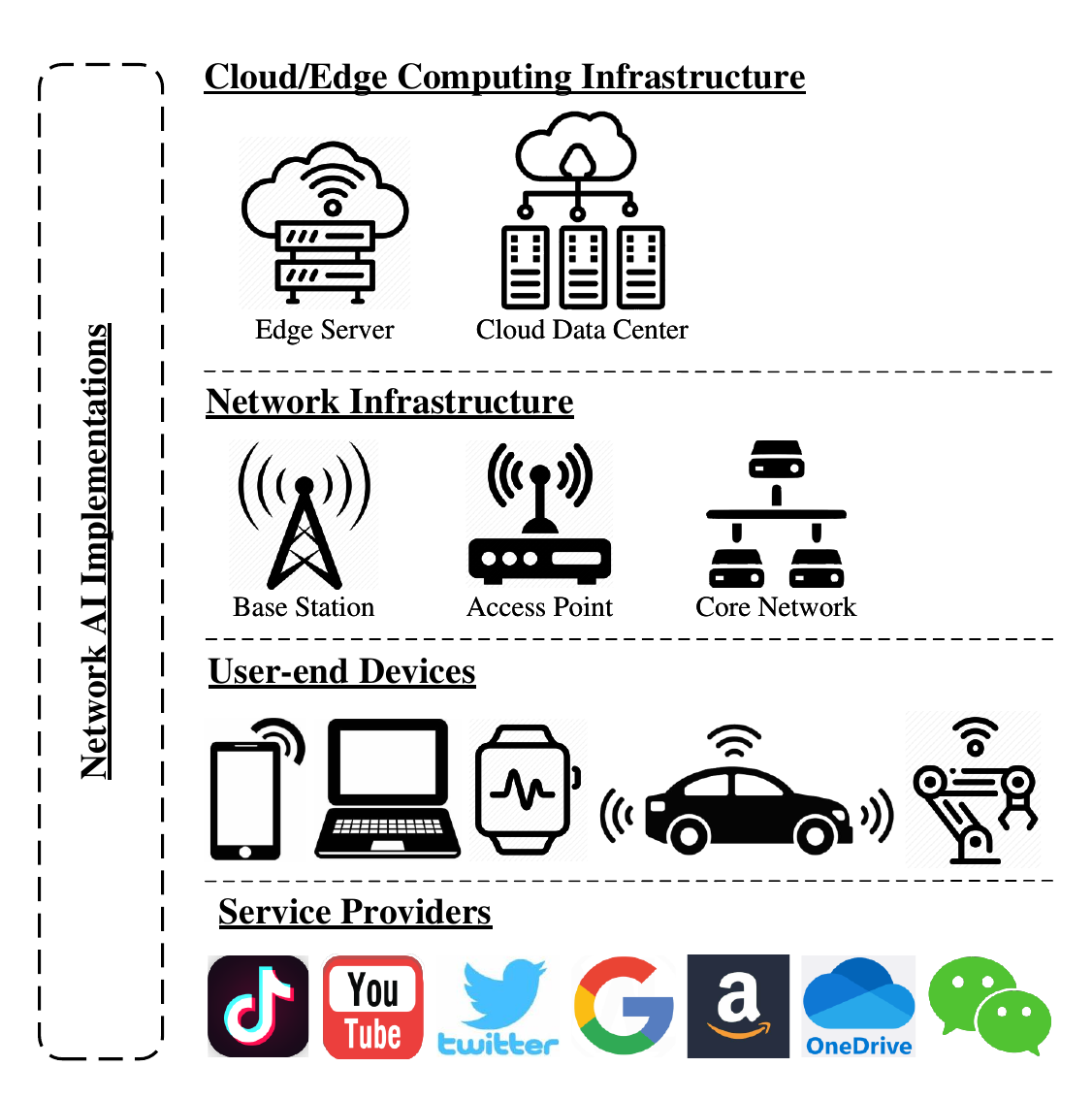}
\caption{Major components and carbon emission sources of a 6G Networking System.}
\label{fig_6Garchitecure}
\end{figure}

Compared to the existing 5G systems, 
major sources of carbon emissions in 6G will exhibit some unique features as illustrated in Fig. \ref{fig_6Garchitecure} and also described in detail as follows:


\noindent{\bf User-end Devices:} 
%
User-end devices need to be charged regularly and, more importantly, they tend to be replaced much more frequent than any other devices or parts of a networking system. This results in a huge amount of carbon emission during the device manufacturing and end-of-life disposal stages. A recent study suggests that, for an average mobile device, the application stage, including daily charging and offering various services, only counts for around 24\% of the carbon footprint throughout its lifecycle, while the rest of carbon footprint is mainly generated during the manufacturing and recycling process\cite{GreenG2022}.  



The next generation networking technology is expected to continuously transform human's daily lives by introducing novel forms of user-end devices including wearables, various forms of IoT devices, connected vehicles and unmanned aerial vehicles (UAVs)\cite{YangYang20226G}. It is expected that the user-end devices will become one of the major sources of carbon emissions in the 6G networking systems. Developing novel solutions to achieve net-zero carbon emission for user-end devices throughout their entire lifecycle will be one of the key targets for future development of networks. 




\noindent{\bf Network Infrastructure:}
%
%
The network infrastructure has long been coined as the major energy consumer, constituting over 70\% of the total energy of a networking system. With the fast rolling out of more advanced networking technology such as 5G-Advanced,  
the carbon emission of the network infrastructure is expected to be further increased in the future. Recent report suggested that, after all the major telecommunication operators are finishing deploying 5G network infrastructure in 2026 in China, the energy consumption of the telecommunication network infrastructure will constitute around 2.1\% of the entire nation's energy supply. Currently, there is still no optimization framework that can balance the demand on the high-performance networking and the requirement on environmental impact minimization.   

The network infrastructure can be further divided into 
radio access network (RAN) 
and core/backbone network. 
%
The energy consumption of RAN accounts for around 50.6\% of the total sum of energy in a standard 5G network. Compared to 4G, there is around 70\% increase in power consumption of RAN in 5G. As an example, a standard 4G base station consumes around 7 kW of power, while a 5G base station will need at least 11 kW of power to operate. For the base stations that carry large antenna arrays, the power consumption could reach over 20 kW. 
%
The core/backbone network consumes around 13.3\% of the total energy of a 5G network.

Recently, some telecommunication operators have announced plans to upgrade their network infrastructures by installing edge computing servers for support AI-based smart services. Also, recent activities in 3GPP, ITU-T, and GSMA have all suggested to employ more AI-based solution for optimizing certain core network functions. This may result in further increase in the carbon emission of network infrastructure in 6G network. It is therefore of critical importance to adopt more lightweight AI models and reuse previously trained AI models and inference results to further reduce the computational and storage need when performing each AI task. 






\noindent{\bf Cloud/Edge Computing Infrastructure: }
The fast growing demand on the cloud-based services significantly increases the energy consumption as well as the corresponding carbon emission of the networking infrastructure for transporting all the data from the users to the cloud. It is estimated that the annual global data center traffic has already reached 20.6 Zettabytes by the end of 2021, over three times of that in 2016. The increasing speed of this number is expected to grow even further in the future. There is a pressing need to develop effective solutions to jointly optimize the environmental impact on the networking infrastructure and the storage and computational facility for achieving net-zero emission in the cloud. 
%
%
Different from the cloud data center, 
edge computing networks  are decentralized in nature and therefore, developing simple decentralized solutions that can allow simple and effective coordination and resource sharing among different edge servers is of critical importance for achieving net-zero edge intelligence-based 6G networks\cite{XY2018TactileInternet}.

\noindent{\bf Service Providers: }
With 5G being transformed from the traditional data-focused communication architecture  to the service-based architecture, evaluating the environmental impact of various services becomes much more important than ever. A traditional metric standardized by ITU-T for measuring the energy efficiency of communication technology is the Energy-Per-Bit, that is the total amount of energy consumed for sending each bit of data. Energy-Per-Bit can be useful when accessing the energy efficiency of data-oriented services. They are however difficult to evaluate the environmental impact of many emerging services with different requirements and design focus. 
Also, 6G is expected to enable many novel services, such as immersive communications and Tactile Internet\cite{XY2018TactileInternet}, driven by various human needs, especially smart services relying on AI-based adaptation and decision making capabilities. There is a pressing need to develop novel metric that is able to evaluate the carbon emission of these emerging services, especially the network AI-enabled services. 

\section{Dynamic Energy and Task Allocation Framework}

The total carbon emissions of a network AI solution is closely related to two important factors: (1) the amount of carbon emitted during the production of the energy to power the networking infrastructure as well as the edge and cloud servers, and (2) the allocation of relevant tasks among different parts of the networking systems as well as the cloud and edge servers.

Motivated by the above observation, we propose a dynamic energy and task allocation optimization framework, referred to as DETA, in which the reallocation of renewable energy sources and distribution of tasks can be jointly optimized with the main objective for minimizing the overall carbon emissions of a network AI implementation throughout a networking system. DETA does not optimize the implemented AI algorithms or software and hardware platforms, but simply redistributes the energy supply and processing servers of different tasks based on the locations, energy and resource types and availability, as well as computational, storage, and transportation costs of task relevant models and data. It can be considered as a combination of the dynamic (harvested renewable) energy trading framework and task offloading solutions for computational and networking converged systems, both have been developed in our previous works\cite{XY2016DETJSAC,XY2018EHFogComputing}.

\noindent{\bf Dynamic Energy Trading (DET):} Some edge/cloud servers have installed energy harvesting devices that can collect energy from their ambient (renewable) energy sources to directly power their operation of computational tasks\cite{XY2018EHFogComputing}. These servers can exchange and share their harvested energy 
with the main objective to minimize the overall carbon emission of the systems. One of the key challenges for DET is that the energy generated by the renewable sources are highly random and can be affected by many uncontrollable and unpredictable influencing factors such as the weather and users' service demands. Therefore, in \cite{XY2016DETJSAC}, we model the dynamic energy trading between different edge servers as a stochastic game in which players (edge servers) that have harvested more renewable energy than they need can negotiate and sell their extra energy to those who cannot obtain sufficient renewable energy to support their local demand. Note that the carbon emission reduction of DET may also be affected by the energy trading loss, e.g., energy  loss during exchange and transportation between edge servers, as well as the storage at various sized battery at each device. 

\noindent{\bf Dynamic Task Allocation (DAT):} In addition to trading their harvested energy, edge servers can also exchange their data communication and processing tasks. In particular, the edge servers that are supplied by high emitting energy sources, e.g., fossil fuel, can offload some of its local data processing tasks to the edge servers supplied by low emitting energy sources, e.g., renewable energy sources.
The high emitting edge servers during this process can turn into sleeping mode for energy saving. The carbon emission reduction achieved by DAT is also closely related to the carbon emission caused by the data transportation between edge servers, as well as the local storage size at each edge server.  

\noindent{\bf Dynamic Energy and Task Allocation (DETA):} In DETA, edge servers jointly optimize the volume of energy as well as the number of data batch exchanged among different edge servers. To simplify our discussion and demonstrate the maximum amount of carbon emission that can be reduced by DETA, we consider the complete information case and assume each edge server can know all the other servers' harvested energy, battery levels, as well as storage information. The incomplete information case can be directly obtained by extending the solutions developed in \cite{XY2016DETJSAC}. Note that DETA is a general carbon emission optimization framework in which various existing optimization methods can be applied according to the specific service requirements.

\section{Case Study: An FEI-based Network AI Implementation}
\label{Section_CaseStudy}

\subsection{Architecture Overview}
\begin{figure}
\centering
\includegraphics[width=9cm]{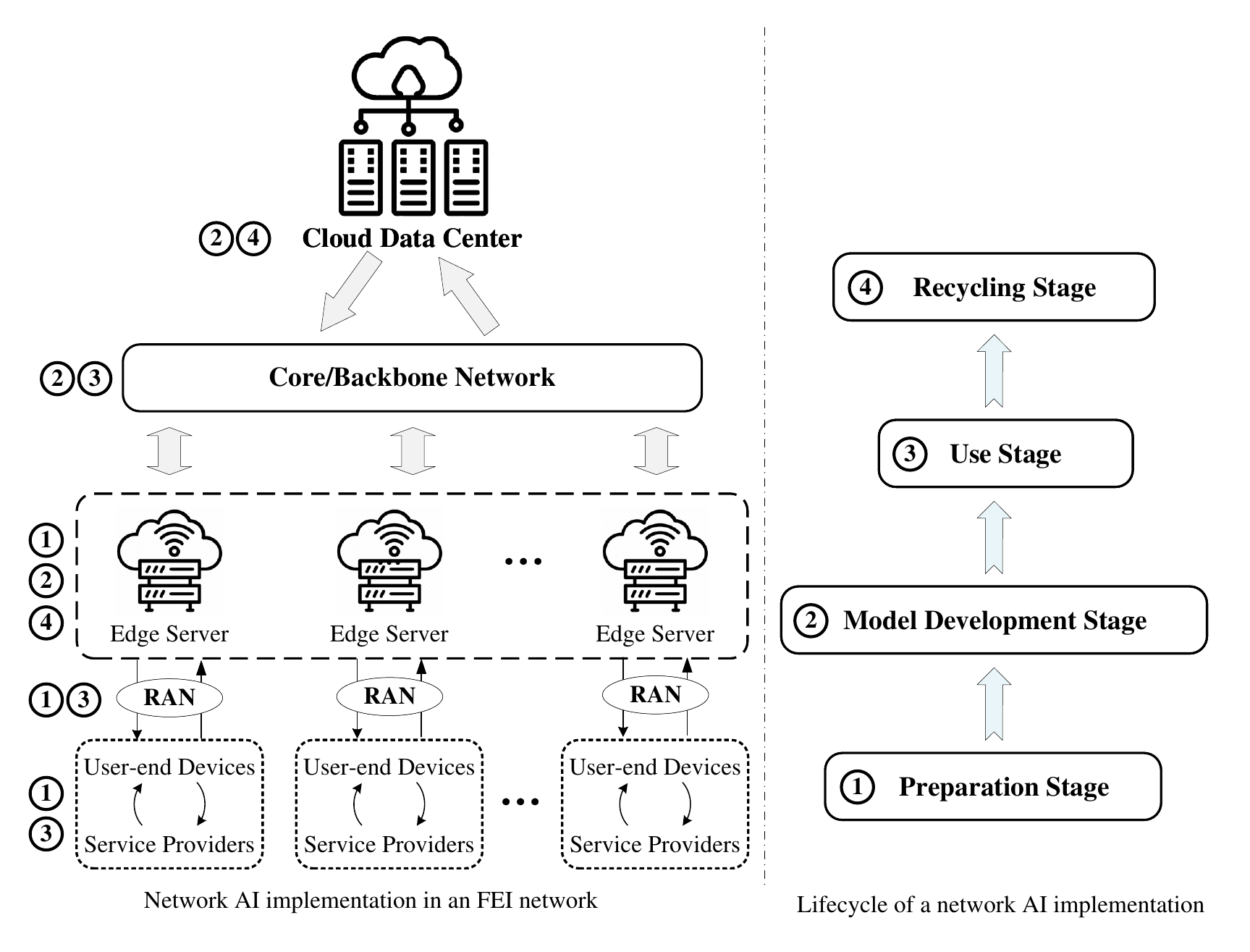}
\caption{An FEI-based network AI implementation in a networking system as well as the corresponding lifecycle of the implementation. }
\label{fig_stages}
\end{figure}

In this section, we investigate the carbon emission of an FEI network, an emerging framework focusing on implementing federated learning-based AI algorithms into the edge computing network\cite{XY2023TimeSensLearning}. FEI has been considered as one of the key enabling technological frameworks to implement network AI in 6G\cite{XY2020Selflearning}. One of the key advantages of FEI is that user-end devices do not have to upload their local raw data to the cloud data center, so the data privacy can be preserved. 
Once received the data samples  from the local user-end devices, each edge server will then train a local model and periodically coordinate with other edge severs through the local model training parameters. An FEI-based network AI implementation architecture and its  implementation lifecycle are illustrated in Fig. \ref{fig_stages}. 

\subsection{Experimental Setup}
\label{Subsection_ExperimentalSetup}
To evaluate the carbon emission of the FEI network, we develop a hardware platform consisting of 10 Raspberry Pi version 4B mini-computers as edge servers, each has a Quad-core Cortex-A72 (ARM V8) 64-bit SoC operated on 1.5GHz. All the edge servers are wirelessly connected to a coordinator (another Raspberry Pi version 4B) via a TP-LINK Wi-Fi Router. To estimate the energy consumption of each edge server during various stages of the network AI implementation, we connect a multi-function USB multi-meter POWER-Z KM001C to the power port of each edge server to measure and keep track of voltage, current, and power during dataset receiving, model training and serving process. We set the sampling rate of each multi-meter to 1 kHz.

We consider two inference tasks and three AI model implementations:

\noindent{\bf Task 1-Image Classification:} We consider the MNIST dataset, a commonly adopted dataset consisting of 60,000 training samples and 10,000 test samples, all are handwritten digit images. We assume all the data samples are uniformly randomly distributed at different edge servers. We set the target accuracy level to 97.5\%. We train the image classification model to differentiate different handwritten digits based on two models: an MLP with 2 hidden layers with 200 units using ReLu activations and a softmax output layer, consisting of 159,010 total parameters and a CNN consisting of two 5$\times$5 convolution layers with 16 and 32 channels, respectively, each followed with 2$\times$2 max pooling, and two fully connected layers with 120 and 84 units, respectively, which has 90,242 parameters in total.

\noindent{\bf Task 2-Next-Character-Prediction:} We consider a language dataset consisting of combinations of words built 
based on William Shakespeare's works. We train an LSTM model based on 3,564,579 training samples (characters) and 870,014 test samples (characters). Our model consists of 2 LSTM layers, each has 256 nodes, and a softmax output layer with one node per character. The model has 866,578 parameters in total.

To calculate the carbon emission of each edge server, we 
use the real carbon emission per kWh data published in 10 different regions across USA\cite{Jacques2019USCarbon} and assume different edge servers in our considered FEI network prototype are randomly distributed across these 10 different regions.

\begin{table}[h!]
  \begin{center}
    \caption{{\footnotesize Energy Consumption and Carbon Emissions of Training Different Models for Different Tasks}}
    \footnotesize
    \label{Table_CompareModels}
    \begin{tabular} {c|c|c|c|c|c|c}
    \hline
      \multirow{2}{*}{Model} & \multirow{2}{*}{Dataset} & \# of & \multirow{2}{*}{$E$} & \multirow{2}{*}{$B$} & Total & CO$_2$e \\
       & & Servers & & & Energy(kWh) & (g)\\
       \hline
       \multirow{6}{*}{MLP}& \multirow{6}{*}{MNIST} & 1 & 50 & 1200 & 0.0034 & 0.08\\
       & & 3 & 50 & 600 & 0.0041 & 0.10\\
       & & 5 & 50 & 300 & 0.0059 & 0.21\\
       & & 7 & 50 & 200 & 0.0074 & 1.11\\
       & & 9 & 50 & 150 & 0.0083 & 1.99\\
       & & 11 & 50 & 120 & 0.0086 & 2.83\\
       \hline
       \multirow{6}{*}{CNN}& \multirow{6}{*}{MNIST} & 1 & 50 & 1200 & 0.0110 & 0.26\\
       & & 3 & 50 & 600 & 0.0120 & 0.29\\
       & & 5 & 50 & 300 & 0.0141 & 0.49\\
       & & 7 & 50 & 200 & 0.0163 & 2.07\\
       & & 9 & 50 & 150 & 0.0196 & 3.93\\
       & & 11 & 50 & 120 & 0.0240 & 6.07\\
       \hline
       \multirow{6}{*}{LSTM}& \multirow{6}{*}{Shakespeare} & 1 & 50 & 300 & 1.8460 & 0.04k\\
       & & 3 & 50 & 300 & 5.0605 & 0.12k\\
       & & 5 & 50 & 300 & 7.6385 & 0.26k\\
       & & 7 & 50 & 300 & 9.5800 & 1.2k\\
       & & 9 & 50 & 300 & 12.0307 & 2.6k\\
       & & 11 & 50 & 300 & 14.0039 & 4.3k\\
       \hline
    \end{tabular}
  \end{center}
\end{table}

\begin{figure}
\centering
\includegraphics[width=8cm]{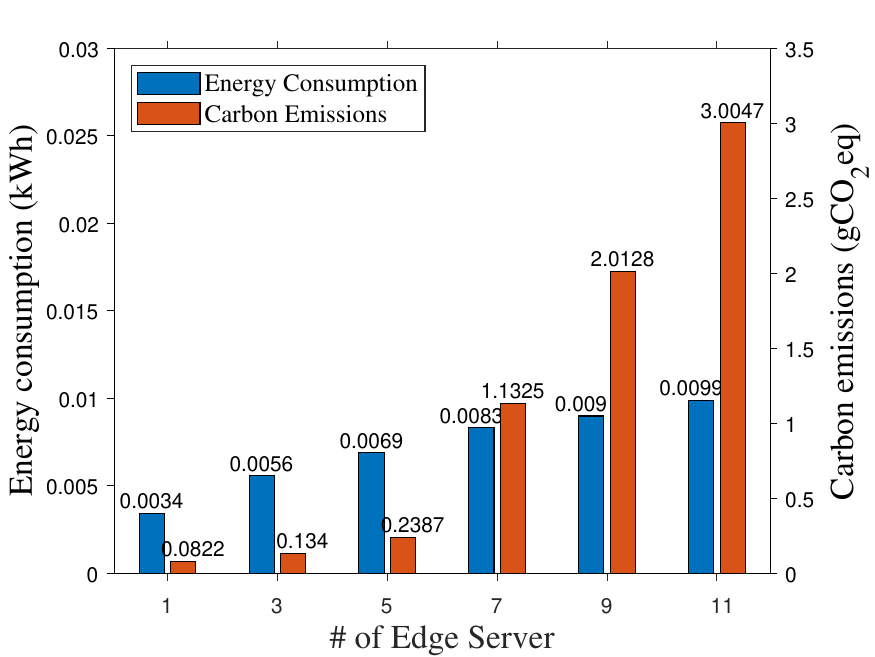}
\caption{Energy consumption and carbon emissions of an FEL network with different numbers of edge servers.}
\label{fig_SimEnergyCarbonFEI}
\end{figure}

\begin{figure}[htbp]
 \centering
   \includegraphics[width=2.5in]{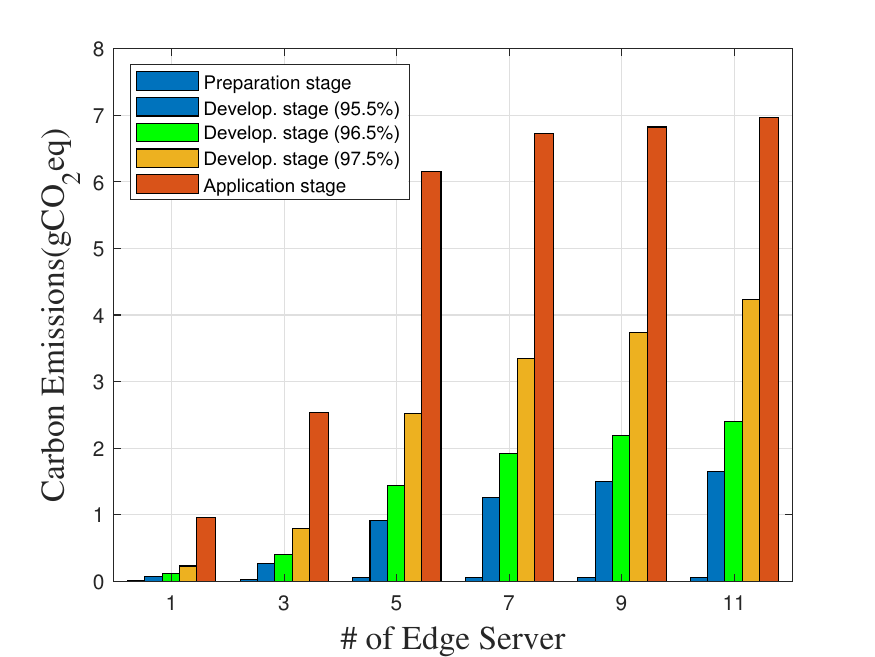}
  \footnotesize
  \caption{\footnotesize{Carbon emissions of an FEL network at different stages of its lifecycle and the portions of different individual stages in the total carbon emissions under different target model accuracy levels. }}
\label{fig_SimLifecycle}
\end{figure}

\begin{figure}
\centering
\includegraphics[width=8cm]{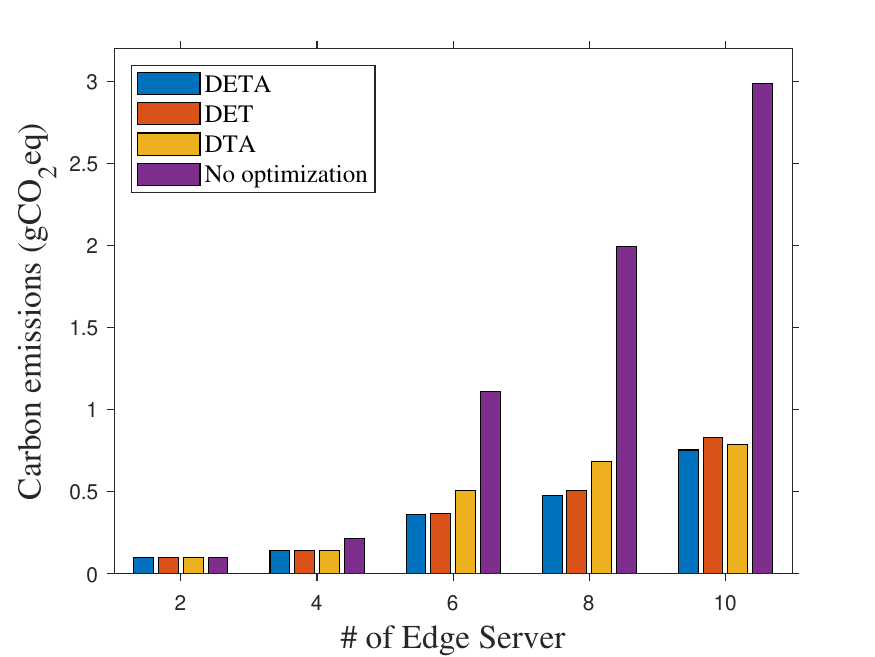}
\caption{Comparison of carbon emissions of network AI implementation  when using different optimization solutions.}
\label{fig_Optimization}
\end{figure}

\subsection{Experimental Results}
In Fig. \ref{fig_SimEnergyCarbonFEI}, we evaluate the overall energy consumption as well as carbon emissions of an FEI network with different numbers of edge servers. We also present the energy consumption when the same AI model is trained in a single edge server deployed at the lowest carbon emission location for comparison (1 edge server case in Fig. \ref{fig_SimEnergyCarbonFEI}). We can observe that, when the energy consumed by the data transportation is negligible, the overall energy consumption of a distributed network AI implementation increases linearly with the total number of edge servers. This is because for any electronic devices involved in the networking system, it will consume static power, also called leakage power, even in the idle state, e.g., an edge server will consume power even when it does not perform any specific task. The more edge servers involved in the network AI implementation, the higher static power consumed by the entire networking system. Since we assume edge servers are randomly located across 10 different considered regions with different carbon emissions per unit of electricity supply. Therefore, the more edge servers being selected to participate in the model training process, the higher the chance that at least one edge server being deployed in the location with high carbon emissions. 
This explain why the total carbon emissions grow in a much faster speed than the energy consumption as the number of edge servers increases.

We present the carbon emissions of a network AI implementation at different stages of its lifecycle in Fig. \ref{fig_SimLifecycle}. We can observe that the carbon emissions of the network AI increase with the total number of edge serves in all three stages. Also, the portions of different stages in the overall carbon emissions depend not only on the number of collaborating edge servers, but also on the model accuracy required by the specific applications, i.e., the higher the required model accuracy, the larger portion of the model development stage.
In Fig. \ref{fig_Optimization}, we compare the carbon emission of the FEI network with and without using our proposed carbon emission optimization solutions including DET, DAT, and DETA. To make our experimental results comparable, we set the maximum portion of tasks offloaded by each edge server to others as well as the maximum percentage of locally harvested renewable energy to be transferred to other edge servers both at 50\%. We can observe that the carbon emission reduction offered by DETA increases with the number of edge servers. When 10 edge servers are participating the network AI implementation, DETA can offer up to 74.9\% reduction in carbon emissions. 

\blu{In Table \ref{Table_CompareModels}, we compare the energy consumption and carbon emissions of training different AI models with different local epoch number $E$ and mini-batch size $B$ for two inference tasks discussed in Section \ref{Subsection_ExperimentalSetup}}. We can observe that even performing the same task with the same target accuracy based on the same dataset, different models can result in different energy consumption, e.g., when performing the image classification task, CNN consumes more energy and generates more carbon emissions than MLP.



\section{Open Problems}

\subsection{Joint Lifecycle Optimization}
The environmental impact of different stages throughout the lifecycle of an AI implementation can be colsely related to each other. For example, the material and manufacturing process at the product stage may directly affect the cost and complexity in the end-of-life recycling and disposal of the device. Evaluating and jointly optimizing the carbon footprint of different stages may involve efforts from various research areas across material sourcing, manufacturing, product shipping and installing, e-waste recycling, etc.

\subsection{New Evaluation Metrics}
As mentioned earlier, traditional metric such as Energy-per-Bit will no longer be able to reflect the environmental impact of the modern mobile services, especially network AI-enabled smart services. This issue is further complicated by the fact that 6G will be focusing more on optimizing and enhancing users quality-of-experience (QoE), which involves subjective human-related factors such as feelings, emotions, and semantics\cite{XY2023iSANJSAC}. How to quantify the impact of QoE  on the service environmental impact is still an important open problem.

\subsection{Low-cost Model Generalization and Transfer}

It is known that the modern AI solutions can excel at some very specific tasks. It is however difficult to generalize or be adopted to other tasks without huge amounts of training. In other words, modifying and transferring AI solutions between different tasks and services can be an expensive and environmental unfriendly process. It is therefore of critical importance to develop novel network AI solutions that are general 
and/or transferable across a wide range of scenarios.


\section{Conclusion}

This article has studied 
the carbon emissions of network AI solutions for 6G and beyond systems. 
An evaluation framework for analyzing the lifecycle of a network AI implementation has been introduced. A novel optimization solution, called DETA, 
has been proposed to optimize the carbon emission of the entire lifecycle of a network AI implementation. 
Case study based on an FEI-based network has been presented. Experimental results have suggested that DETA has the potential to significantly reduce the carbon emission in future AI-native networking systems.

\section{Carbon Impact Statement}
The experiment of this article consumed 50.29 kWh of electricity, which is equivalent to 29.27 kg CO$_2$eq of carbon emissions based on the Carbon intensity reported by ``2021 Carbon intensity report for industrial electricity in China". 

\section*{Acknowledgment}
The work of Yong Xiao was supported in part by the National Natural Science Foundation of China (NSFC) under Grant 62071193 and in part by the Major Key Project of Peng Cheng Laboratory under Grant PCL2023AS1-2. The work of Yingyu Li was supported in part by the NSFC under Grant 62301516. The work of Guangming Shi was supported in part by the NSFC under Grant 62293483. The work of Xiaohu Ge was supported in part by the NSFC under Grant U2001210. This work of Yang Yang was partially supported by the National Key Research and Development Program of China under Grant 2020YFB2104300 and the NSFC under Grant U21B2002.


\bibliographystyle{IEEEtran}
\bibliography{DeepLearningRef}

\begin{IEEEbiographynophoto}{Peng Zhang} 
is currently pursuing his PhD in the school of electronic information and communications at the Huazhong University of Science and Technology, Wuhan, China. He is also a senior engineer at the FiberHome Telecommunication Technologies Co. Ltd., His research interest includes network AI, edge intelligence, and broadband network.
\end{IEEEbiographynophoto}

\begin{IEEEbiographynophoto}{Yong Xiao}(Senior Member, IEEE) is a professor in the School of Electronic Information and Communications at the Huazhong University of Science and Technology, Wuhan, China. He is also with the Peng Cheng Laboratory, Guangzhou, China and the Pazhou Laboratory (Huangpu), Guangzhou, China. 
His research interests include 
semantic communication, cloud/fog/edge computing, green networks, and IoT.
\end{IEEEbiographynophoto}

\begin{IEEEbiographynophoto}{Yingyu Li} (Member, IEEE) is an associate professor in the School of Mechanical Engineering and Electronic Informaton at the China University of Geosciences (Wuhan). 
Her research interests include  distributed optimization and learning, and their applications in edge computing, green communication systems, and IoT.
\end{IEEEbiographynophoto}

\begin{IEEEbiographynophoto}{Xiaohu Ge} (Senior Member, IEEE) is currently a Professor with the School of Electronic Information and Communications at Huazhong University of Science and Technology, China. 
His research interests are in the area of mobile communications, traffic modeling in wireless networks, green communications, and interference modeling in wireless communications.
\end{IEEEbiographynophoto}

\begin{IEEEbiographynophoto}{Guangming Shi} (Fellow, IEEE) is 
now the deputy director of Peng Cheng Laboratory, Shenzhen, China. He is also a Professor with the School of Artificial Intelligence, Xidian University. 
His research interest includes semantic communication, brain-inspired computing, and signal processing.  
\end{IEEEbiographynophoto}
%
%
\begin{IEEEbiographynophoto}{Yang Yang}(Fellow, IEEE) is a professor with the IoT Thrust at the Hong Kong University of Science and Technology (Guangzhou), China. He is also the Chief Scientist of IoT at Terminus Group, an adjunct professor with the Peng Cheng Laboratory, and a Senior Consultant for Shenzhen Smart City Technology Development Group, China. His research interests include multi-tier computing networks, 5G/6G systems, and AIoT technologies and applications.

\end{IEEEbiographynophoto}
%
%

\end{document}